%% file: main.tex
\documentclass[sn-nature]{sn-jnl}

\input{sections/preambles}

\begin{document}

\title[title]{AI-aided Geometric Design of Anti-infection Catheters}

\author[1,2]{\fnm{Tingtao } \sur{Zhou}}
\equalcont{These authors contributed equally to this work.}

\author[3]{\fnm{Xuan} \sur{Wan}}
\equalcont{These authors contributed equally to this work.}

\author[1]{\fnm{Daniel Zhengyu} \sur{Huang}}

\author[1]{\fnm{Zongyi} \sur{Li}}

\author[2]{\fnm{Zhiwei} \sur{Peng}}

\author[1]{\fnm{Anima} \sur{Anandkumar}}

\author[1,2]{\fnm{John F.} \sur{Brady}}

\author*[3]{\fnm{Paul W. } \sur{Sternberg}}
\email{pws@caltech.edu}

\author*[1]{\fnm{Chiara } \sur{Daraio}}
\email{daraio@caltech.edu}

\affil[1]{\orgdiv{Division of Engineering and Applied Science}}

\affil[2]{\orgdiv{Division of Chemistry and Chemical Engineering}}

\affil[3]{\orgdiv{Division of Biology and Biological Engineering}}

\affil{\orgname{California Institute of Technology}, \orgaddress{\city{Pasadena}, \postcode{91125}, \state{CA}, \country{USA}}}


\input{sections/0-abstract}

\maketitle

\input{sections/1-main-body}

\backmatter

\input{sections/S2-Acknowledgments}

\bibliography{sn-bibliography}

\newpage

\input{sections/3-methods}

\newpage

\input{supp}


\end{document}

%% file: sections/preambles.tex
\usepackage{graphicx}%
\usepackage{multirow}%
\usepackage{amsmath,amssymb,amsfonts}%
\usepackage{amsthm}%
\usepackage{mathrsfs}%
\usepackage[title]{appendix}%
\usepackage{xcolor}%
\usepackage{textcomp}%
\usepackage{manyfoot}%
\usepackage{booktabs}%
\usepackage{algorithm}%
\usepackage{algorithmicx}%
\usepackage{algpseudocode}%
\usepackage{listings}%

\theoremstyle{thmstyleone}%
\theoremstyle{thmstyletwo}%

\theoremstyle{thmstylethree}%
\newcommand{\revise}{}

%% file: sections/0-abstract.tex
\abstract{

Bacteria can swim upstream due to hydrodynamic interactions with the fluid flow in a narrow tube~\cite{RN17,RN23,RN20,RN19,RN100,kaya2012direct,zottl2012nonlinear,marcos2012bacterial, shen2012flow}, 
and pose a clinical threat of urinary tract infection to patients implanted with catheters
~\cite{RN4, RN3, RN6, RN7, RN2}. 
Coatings and structured surfaces have been proposed as a way to suppress bacterial contamination in catheters.
However, there is no surface structuring or coating approach to date that thoroughly addresses the contamination problem. 
Here, based on the physical mechanism of upstream swimming, we propose a novel geometric design, optimized by an AI model predicting in-flow bacterial dynamics. The AI method, based on Fourier neural operator, offers significant speedups over traditional simulation methods.  Using {\it Escherichia coli}, we demonstrate the anti-infection mechanism in quasi-2D micro-fluidic experiments and evaluate the effectiveness of the design in 3D-printed prototype catheters under clinical flow rates. Our catheter design shows 1-2 orders of magnitude improved suppression of bacterial contamination at the upstream end of the catheter, potentially prolonging the in-dwelling time for catheter use and reducing the overall risk of catheter-associated urinary tract infections.

}

%% file: sections/1-main-body.tex
Catheter-associated urinary tract infections \cite{RN4, RN3, RN6, RN7, RN2} (CAUTI) are among the most common infections in hospitalized patients, costing about 30 million US dollars annually \cite{RN8}. From a materials/device perspective, previous methods to prevent such infections included catheter impregnation with antimicrobial silver nano-particles \cite{RN5}, or the use of antibiotic lock solutions, anti-adhesion, or antimicrobial materials \cite{RN11, RN13}. However, none of these methods surpasses the effect of  stricter nursing procedures, and current clinical practice to prevent CAUTI focuses on reducing catheter in-dwelling time to prevent CAUTI. 
The design of catheters that reduce bacteria motility in the presence of fluids would offer a significant improvement to the state of the art management of CAUTI.

For this, it is crucial to understand the locomotion patterns of microbes in the presence of fluid flow under confinement. A typical microbial trajectory alternates between periods of running (propelling themselves in a straight line) and tumbling (randomly changing direction) to explore the environment \cite{RN40,RN22,RN39,RN23}. Hydrodynamic interactions and quorum sensing lead to more complicated dynamics, such as enhanced attraction to the surface \cite{RN24, lauga2009hydrodynamics}, and collective swarming motion \cite{RN25, kaiser2007bacterial, verstraeten2008living,ghosh2022cross}. Recent studies show that, in shear flows,  microscopic run-and-tumble motion can lead to macroscopic upstream swimming \cite{RN17,RN23,RN20,RN19,RN100,kaya2012direct,zottl2012nonlinear,marcos2012bacterial,shen2012flow}. Normally, passive particles are convected downstream in addition to diffusive spreading \cite{RN97}. However, the self-propulsion of microbes results in qualitatively different macroscopic transport: the body of a bacterium crossing the tract is rotated by fluid vorticity. As a result, near the wall where flow is slow, its head direction is more likely to point upstream, and they swim  upstream along the wall. 
Recent experiments \cite{RN15}  have demonstrated super-contamination of \revise{{\it E. coli}} in a micro-fluidic channel, highlighting the importance of their power-law runtime distribution, which dramatically enhances the tendency of bacteria to swim  upstream, and the bacteria can swim  persistently against the flow.
  
Mainstream strategies to prevent bacterial contamination include: (1) Physical barriers, such as filters or membranes~\cite{RN48, RN46, RN47}; (2) Antimicrobial agents, such as antibiotics \cite{RN50, RN49}; (3) Surface modifications of medical equipment to reduce bacterial adhesion and biofilm formation \cite{RN61, RN62, RN63, RN64,RN65, RN66,RN67}; (4) Control of physical/chemical environment, such as high/low temperatures, low oxygen levels, or disinfectants to suppress bacterial growth and survival \cite{RN51, RN52, RN53, RN54}; (5) Strict sterilization procedures, such as gloving and gowning \cite{RN55, RN56, RN57}; (6) Regularly monitoring patient conditions to detect and treat bacterial contamination early on \cite{RN58,RN59, RN60}. 
Although various surface modifications or coatings have been proposed to reduce bacterial adhesion, none has been shown to prevent upstream swimming or catheter contamination effectively \cite{RN65, RN66,RN67}. Other passive anti-bacterial methods, such as membranes or filtration, may be difficult to apply directly to patients with indwelling catheters. 

Geometric control of microbial distribution is safer than antibiotics or other chemical methods regarding antibiotic resistance \cite{RN68,RN69,RN70,RN71}.
Specific shapes have been used in other contexts to confine and trap undesirable bacteria~\cite{RN26}. \revise{Asymmetric shapes can also} bias the partitioning of motile bacteria~\cite{RN32} due to the “dry” geometric rectification effect, and irregular boundary shapes can create vortices proportional to the curvature.

We sought to engineer catheters that prevent bacteria from swimming upstream and minimize contamination. To optimize the geometry of the catheters, we constrain\revise{ed} the design space to placing triangular obstacles on  the interior catheter's wall. Capturing the physical mechanism of upstream swimming, we perform\revise{ed} coupled fluid particle simulations to discover geometric design principles. We  model\revise{ed} the \revise{bacteria distribution}, by coupling the hydrodynamics and geometric
rectification effects as a stochastic partial differential equation (SPDE). We then use\revise{d} the simulation data to train an AI model based on Geo-FNO,~\cite{RN34, RN35}, to learn the solutions of the SPDE
and use the trained model to optimize the catheter geometry. 
Based on the optimized design, we fabricate\revise{d} quasi-2D microfluidic devices and 3D print prototype catheters, to evaluate the effectiveness of our concept. Our experimental results show up to 2 orders of magnitude improved suppression of bacterial super-contamination compared to standard catheters, suggesting a pathway for the management of CAUTI.

\begin{figure}[h]
\centering
\includegraphics[width=0.9\textwidth]{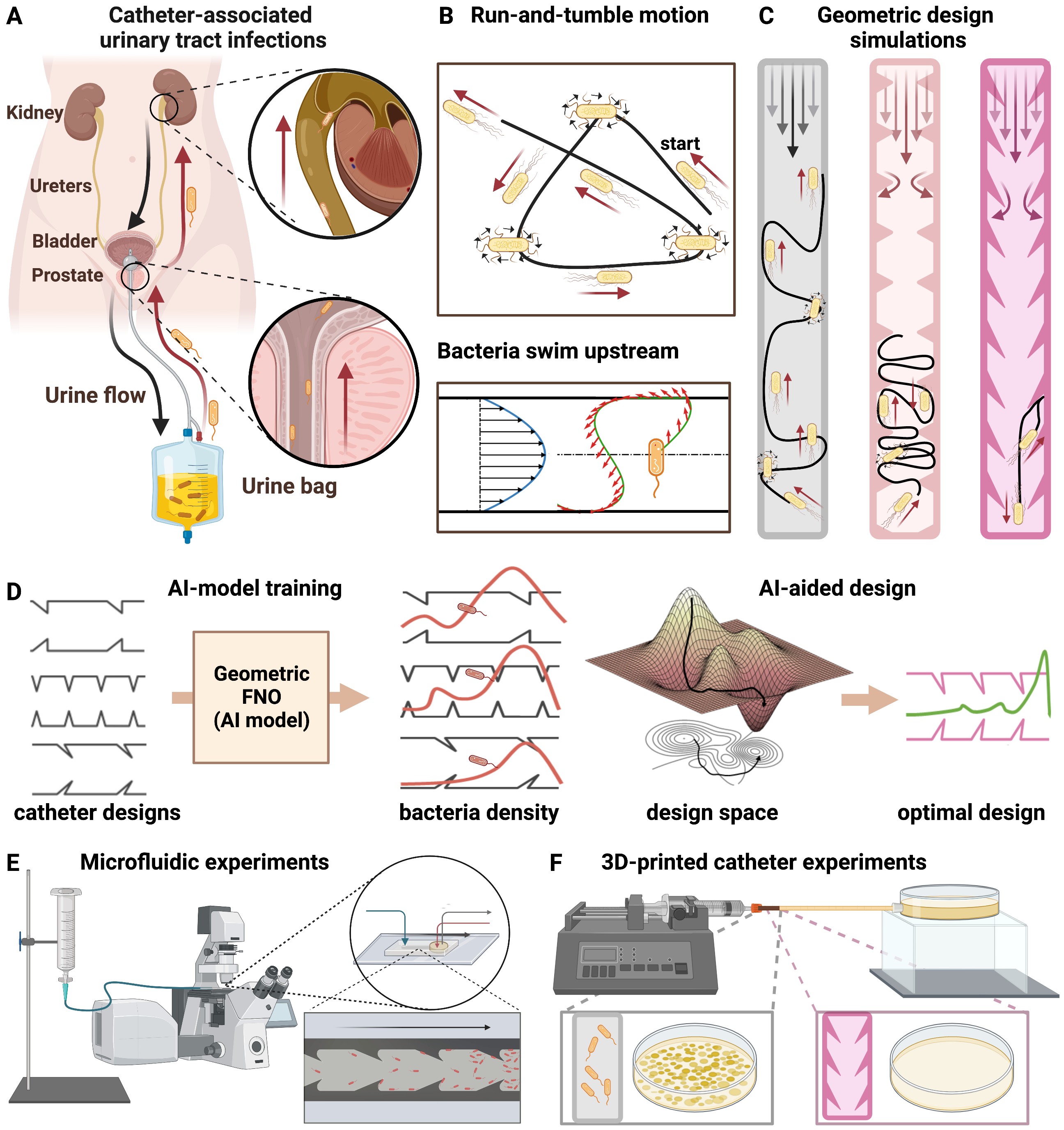}
\caption{(A) Proposed mechanism for catheter-associated urinary tract infections (CAUTI). The urine flows from within the patient’s bladder outward through a catheter, while bacteria swim upstream into the patient's body. (B) The run-and-tumble motion of bacteria and upstream swimming mechanism. (C) Simulations to explore catheter shapes suppressing upstream swimming. (D) AI-assisted optimization using the geo-FNO framework. (E) Microfluidic experiments to test the design in 2D channels. (F) 3D experiment with designed real-size catheters.}
\label{fig1}
\end{figure}

\paragraph{Investigating the Microscopic Mechanism} 
We first stud\revise{ied} numerically the  role of conventional surface modifications, like antimicrobial nanoparticle coatings ~\cite{RN49,RN64}, engineered roughness or hydrophobicity~\cite{goodman2013future,jaggessar2017bio}, in the suppression of bacteria’s upstream swimming. To model their presence,  we assume\revise{d} that they cause the bacteria to detach from the surfaces and dwell at a distance of at least 3 $\mu$m away from the surface, which is above the typical body length of \revise{{\it E. coli}} (1-2 $\mu$m). We find that the upstream swimming behavior is not much affected by surface repulsion at all flow rates tested in our simulations. Comparing the simulated trajectories of a persistent bacterium inside a smooth channel (Fig.  2D) and a surface-modified channel (Fig.  2E), the upstream swimming behaviors are similar. We quantify the effectiveness for the suppression of bacterial upstream swimming by two population statistics: 1. The averaged upstream swimming distance 
$\langle x_{up} \rangle = -\int_0^{-\infty} \rho(x) x \mathrm{d}x$; where $\rho(x)$ is the bacteria \revise{distribution function}. 2. The distance that the top 1\% upstream swimmers arrive at $x_{1\%}$. The simulated surface modification only slightly reduces $\langle x_{up} \rangle$ at intermediate flow rates but barely changes $x_{1\%}$ (blue and pink lines in Fig.  2F). The poor effectiveness of surface modifications is consistent with current experimental observations \cite{RN65, RN66}.

We then explore\revise{d} the role of catheters' surface geometry by adding physical obstacles. We find that symmetric and asymmetric obstacles significantly suppress upstream swimming (black and green lines in Fig.  2F). We identify two synergistic effects: First, the slope of the obstacles redirects the bacteria's swimming direction when they take off from the top of the obstacle, interrupting the continuous climbing along the wall's surface. Asymmetric shapes bias bacteria motion downstream (Fig.   2A),  as shown by simulated trajectories at 0 flow rate (SI and Fig.  S1), and the difference in the upstream swimming statistics (black and green lines of Fig. 2F) at low flow rates. 

Second, at a finite flow rate, the flow field differs from the Poiseuille flow in a smooth channel (Fig.  2B). In the Poiseuille flow, the vorticity turns the bacteria downstream. Near the top of the obstacle, the flow speed and vorticity are greatly enhanced (Fig. 2C and Fig. S2), boosting the turning mechanism. Combining these two effects, we expect upstream swimming to be 
significantly reduced in channels with optimized obstacles' geometry.

\begin{figure}[h]%
\centering
\includegraphics[width=0.9\textwidth]{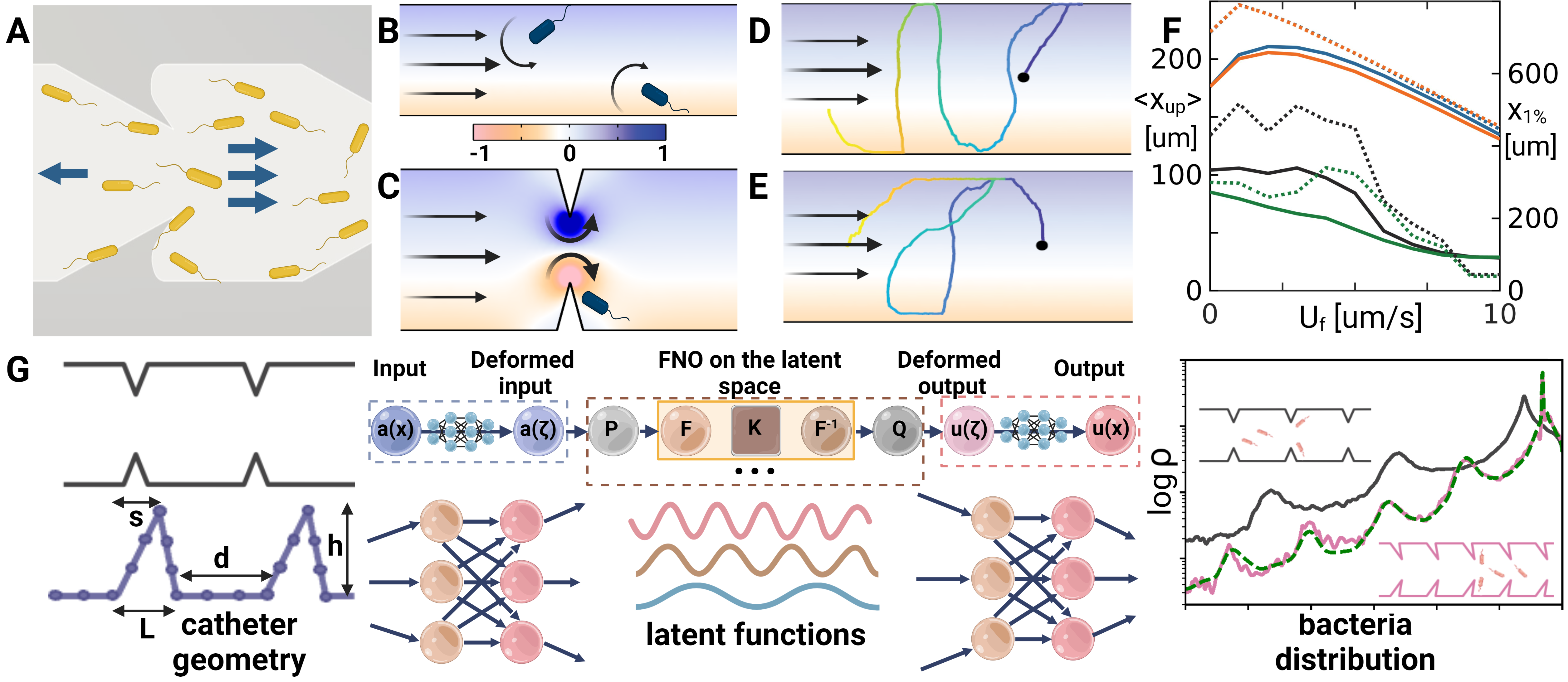}
\caption{Physical mechanism of obstacle suppressing upstream swimming and geometric optimization. (A) Geometric rectification effect without flow. (B-E): Color background shows  the relative magnitude of flow vorticity (darker is larger; blue counterclockwise, yellow is clockwise). (B) Poiseuille flow in a smooth channel. The vorticity rotates the head of the bacteria downstream. (C) Flow in a channel with symmetric obstacles. Flow speed and vorticity are enhanced near the top of the obstacle, leading to stronger torque redirecting the bacteria downstream. (D, E) Simulated trajectories of a persistent bacterium in 2D channels of width 50 $\mu$m under different conditions: (D) a smooth channel, (E) a surface-modified channel that repels bacteria. 
(F) Population statistics of upstream swimming. Solid lines (left y-axis) show the average upstream distance. Dashed lines (right y-axis) show the upstream distance of the top 1\% swimmers in the population. \revise{Blue lines for smooth channels, orange for surface-modified channels, black for symmetric obstacles, and green for asymmetric obstacles.} 
(G) Design optimization.
The AI model maps the catheter interior shape to the bacteria \revise{distribution} in the forward simulation and inversely optimizes the geometric parameters.
Inset on the right panel shows the random initial condition (black) and the optimized design (pink). The optimized \revise{bacteria distribution} is verified by simulation (green dashed line).
}
\label{fig2}
\end{figure}

\paragraph{AI-aided Optimization of Geometric Conditions}

We use\revise{d} \revise{an} AI model to optimize  the channel shape, characterized by four parameters: obstacle base length $L$,  height $h$, tip position $s$, and inter-obstacle distance $d$ (Fig. 2G). This method first maps the irregular channel geometry to a function in the latent space (a unit segment [0,1]), then applies the FNO model in the latent space, and finally transforms the \revise{bacteria distribution} back to the physical space (Fig. 2G). 
We then use\revise{d} this trained surrogate model for inverse design optimization, to determine the optimal channel shape. To evaluate the effectiveness of each design, we measure the averaged $\langle x_{up} \rangle$ at $T=500s$ for three flow \revise{speeds} (5, 10, 15 $\mu$m/s).
Our AI-aided shape design, based on geometry-aware Fourier neural operator, outperforms given shapes in training data by about 20\% in terms of weighted \revise{bacteria distribution}. It is also extremely fast compared to traditional numerical simulations – it took 15 seconds for our trained AI model to generate the optimal design vs. three days with a traditional solver. Even after accounting for the time it takes to generate training data (10 minutes) and the time to train the model (20 minutes), we achieve a significant speedup compared to using a traditional solver for design optimization.

\paragraph{Microfluidic Experiments} To evaluate the effectiveness of the optimized structure, we fabricated quasi-2D microfluidic channels to observe bacterial movement under a microscope (Fig. 3A). We select\revise{ed} the subset of upstream swimming bacteria and categorize\revise{d} them according to where they detach from the walls. A trajectory is denoted as ‘type 1’ if it detaches from the top of the obstacle (Fig. 3D upper); ‘type 2’ if it detaches from the smooth part of the wall (Fig.  3D lower). Between 70-80\% of upstream swimming trajectories belong to type 1, for flow rate $U_0<100 \mu m/s$ (Fig. 3E). We also notice\revise{d} that all the upstream swimming trajectories observed from these experiments are redirected downstream (Fig.  3E red line). 
Bacteria accumulation was observed near a sharp corner (Fig. 3B), possibly due to the stagnation zone (Fig.  2C and Fig. S1 white color near the corner). To prevent bacteria accumulation at the corners, we round\revise{ed} the geometry with an arc of radius $r=h/2$ (Fig. 3C).

\begin{figure}[h]%
\centering
\includegraphics[width=0.9\textwidth]{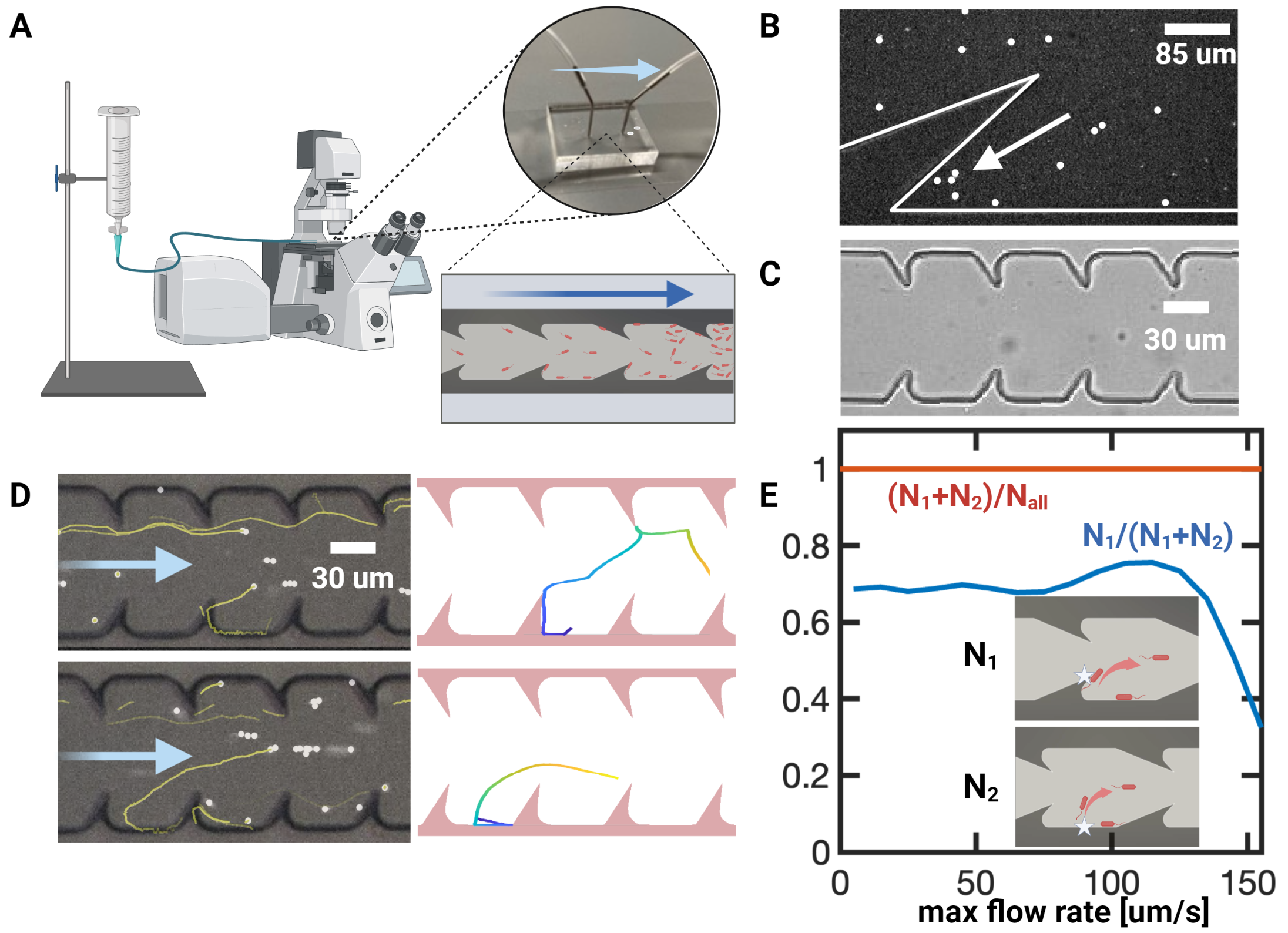}
\caption{Microfluidic experiments: A. Schematic of the microfluidic experiments. One end of the microfluidic  channel is connected to a syringe filled with imaging solution, while the other end is connected to a reservoir of \revise{{\it E. coli}}. The long arrow denotes the flow direction. B.  Bacteria accumulation at the sharp corner due to flow stagnation. C. Bright field image of the microfluidic channel. D. Typical events of bacteria (white dots) falling off the channel walls, with their trajectories of the past 5 seconds shown in yellow lines. The upper image shows a “type 1” trajectory where the bacteria falls off from the obstacle tip. The lower image shows a typical “type 2” trajectory where the bacteria falls off from the smooth part of the wall. Left column experimental; right column simulation. E.  Statistics of fall-off events.}
\label{fig3}
\end{figure}

\paragraph{Macro-scale Catheter Experiments} The mechanisms and design principles demonstrated above are readily scaled up to  catheters. In 3D tubes, bacteria can traverse the tube through any cutting line of the cross-section. Bacteria moving near the boundary can still swim upstream due to the same mechanism shown above (Fig. 2 A,B,F-I and Fig. S1),  where only the dimensionless shear rate near the wall matters \cite{RN20}. The run length of  supercontaminating bacteria  can exceed 1 mm \cite{RN15}, comparable to the rescaled obstacle size, and the rectification effect is expected to persist at these scales\cite{RN32}. The linear scaling of the Stokes flow also guarantees that the round-off of the corners will eliminate flow stagnation zones.  
We 3D print\revise{ed} prototype catheters to test design effectiveness at medically relevant spatial and temporal scales. For these, we enlarge\revise{d} the obstacle's size to scale for a catheter with a 1.6 mm inner diameter and revolve their geometry around the center line. The obstacles become extruded rings, 0.4~$\mu m$ tall, with 1 mm spacing. Typical human urine rate is estimated to be $25-60 mL/hr$ \cite{chenitz2012decreased,macedo2011defining,kellum2015classifying} for adults, and lower for children. Since upstream contamination is stronger at lower flow rates, we test between 0 and $10 mL/hr$. One end of the tube is connected to a syringe controlled by a mechanical pump, and the other is connected to \revise{{\it E. coli}} reservoir (Fig. 4A). After 1 hr, the tube is cut into 2 cm long segments, discarding the two ends segments, and the liquid inside is transferred to culturing plates. We quantified the \revise{distribution} of bacteria in each segment by counting colonies after culturing the plates for 24 hrs at room temperature (SI and Fig. ~S5). We observed upstream contamination in smooth tubes (16 cm long) at various flow rates ($0, 2.0, 8.9 mL/hr$;  Fig.  4B \revise{and Fig.  S4}). With bacteria velocity of 20 um/s, the most persistent ones reach about 7 cm upstream in 1 hr. As a result, most bacteria concentrate in the two segments near the reservoir. We then tested the difference between smooth and designed tubes (8 cm long). The designed tubes show 1-2 orders of magnitude fewer bacteria contamination compared to the smooth tubes at the same positions (Fig.  4C). At $0 mL/hr$, the suppression is about ten times, due to the geometric rectification effect (Fig.  2A). At $8.9 mL/hr$, we observed more than 100 fold suppression, due to the additional vortex-redirecting effect (Fig. ~2B).
 
\begin{figure}[h]%
\centering
\includegraphics[width=0.9\textwidth]{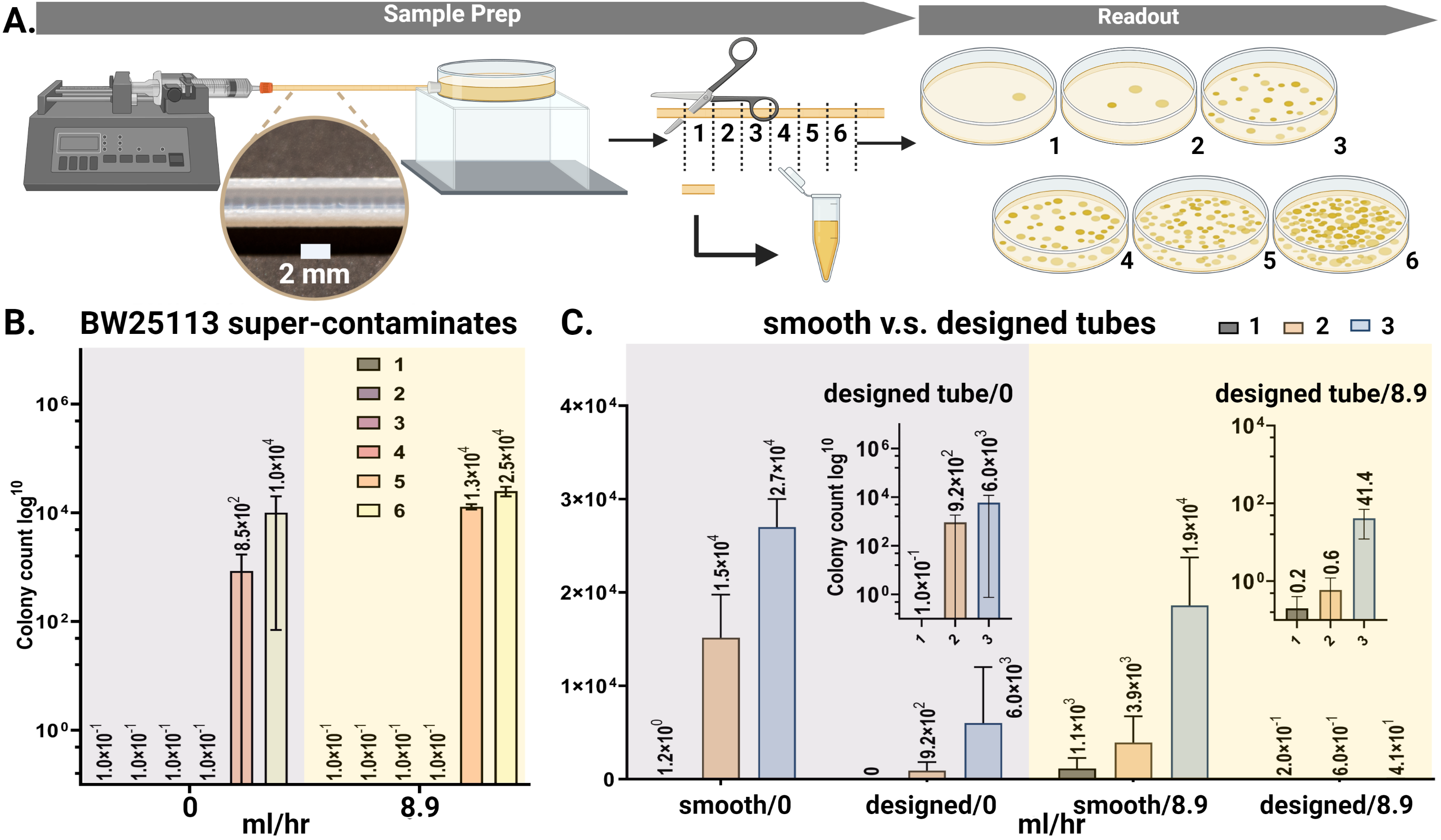}
\caption{Experiments on 3D-printed catheter prototypes. (A) Experimental setup. The downstream end of the tube was connected to a reservoir of {\it \revise{{\it E. coli}}}, and the upstream end was connected to a syringe full of culture solution controlled by a syringe pump. After 1 hr, the tube was cut into equally long segments, and the liquid inside was extracted to culture for 24 hrs. The number of \revise{{\it E. coli}} colonies was counted under a microscope to reflect the amount of bacteria in each segment. (B) Super-contamination of \revise{{\it E. coli}} in smooth tubes. (C) Comparison of designed v.s. smooth tubes.}
\label{fig4}
\end{figure}

\paragraph{Discussion}
In this work, we introduce an effective geometrical design of the interior surfaces of medical catheters for suppressing bacterial upstream swimming and super-contamination. 

Our design approach is based on impeding the physical mechanism for bacterial upstream swimming, considering the general model of spherical particle rheotaxis with power-law dynamics. Details of the bacterial locomotion, such as their flagella chirality~\cite{RN99} and hydrodynamic interaction with the boundary~\cite{RN100}, may modify the simulation results quantitatively regarding specific microbial species, which are neglected here. 
We find a lower bound on the separation between obstacles to maximize the effective vorticity near the tip of obstacles, due to the interaction of overlapping vortices (Fig. ~S2 and Supplementary Discussions). The constraint on obstacle heights is a trade-off between enhancing effective vorticity and avoiding clogging of the tube (Fig. ~S2).

We note that the geometrical design cannot completely eliminate bacterial upstream swimming, especially at near-zero flow rates. However, it drastically reduces the amount of super-contamination and may significantly prolong the indwelling time of catheters.
Using our designed catheters is not expected to require changes to the regular clinical protocols, or retraining of medical personnel. 
Moreover, our solution does not introduce chemicals into the catheters, and thus is safe and does not require additional maintenance. Our geometrical design approach is expected to be compatible with other procedural measures, anti-bacterial surface modification, and environmental control methods. 

%% file: sections/S2-Acknowledgments.tex
\bmhead{Acknowledgments}

We thank J. P. Marken and R. Murray for providing the bacteria strains. We thank D. J. Anderson for providing the microscope used in this study. We thank P. Arakelian for assistance in 3D printing. We thank A. Ghaffari for training on microfluidic fabrication. T.Z. and X.W. thank M. Chen and W. Miao for discussions.  C.D. and J.F.B. acknowledge financial support by the  Donna and Benjamin M. Rosen Bioengineering Center Pilot Research Grant. C.D. acknowledges support from the Heritage Medical Institute at Caltech. D.Z.H. is supported by the generosity of Eric and Wendy Schmidt by recommendation of the Schmidt Futures program. Z.L. is supported in part by the PIMCO Fellowship and Amazon AI4Science Fellowship. A. A. and P.W.S. are supported by Bren Professorships. 
\\
\\
{\bf Author contributions}:
T.Z. and X.W. contributed equally to this work as co-first authors. X.W., T.Z., P.W.S. and C.D. designed experiments. X.W. and T.Z. performed experiments and analyzed data. T.Z. and Z.P. performed simulations. D.Z.H. and Z.L. designed the AI model and performed optimization. 
A.A. conceptualized and planned the AI framework.
T.Z., J.F.B. and C.D. conceived the project. P.W.S. and C.D. supervised the project. All authors discussed the results and contributed to the manuscript writing.\\
\\
{\bf Competing interests}:
California Institute of Technology (Caltech) has a patent pending related to the discoveries in this manuscript. 

Supplementary Information is available for this paper. 
\\
\\
{\bf Materials and Correspondence}:
Correspondence and material requests should be addressed to Chiara Daraio and Paul W. Sternberg.

%% file: sections/3-methods.tex
\section*{Methods}\label{sec11}
\paragraph{Coupled fluid and particle dynamics simulations} 
We simulate\revise{d} the Stokes flow inside a channel with no-slip boundary conditions using the COMSOL software. The resulting velocity and vorticity fields are then coupled into the particle dynamics simulations, while the feedback of particle motion on the fluid dynamics is neglected in the limit of dilute suspension\revise{s} and small particle sizes. The particle dynamics is described by the Active Brownian Particle (ABP) model with Gaussian statistics and the run-and-tumble (RTP) model with power-law (Levy) statistics. The coupled simulations \revise{were} performed with our in-house GPU Julia code with a simulation timestep of $10^{-4}$ s. In the ABP model, individual particle dynamic\revise{s} is integrated according to the over-damped Langevin equation \cite{RN20}
\[0=-\zeta (U-u) + \zeta U_0 q(t) + F(t)\]
$dq/dt=(1/2 \omega+L/\zeta_R )\times q$\revise{, where} $\zeta$ is the viscous drag coefficient, $U$ the particle’s velocity, $q$ the particle’s orientation vector, $u$ the \revise{local} flow velocity, $\omega$ the \revise{local} flow vorticity. $F(t)$ is Gaussian random force satisfying $\langle F(t)\rangle=0$ and $\langle F(0)F(t)\rangle=\revise{2}k_B T\delta(t)I$. $L$ is Gaussian random torque with $\langle L(t)\rangle=0$ and $\langle L(0)L(t)\rangle=\revise{2}\zeta^2 \delta(t)I/\tau_R$, and $\tau_R$ is the average runtime. In the RTP model, individual particles will be displaced with $L(t)=0$ (the `run' phase) for $0<t<\tau_R$. Then $q$ is changed instantaneously to a random new direction (the `tumble') $q'$ and the process \revise{repeated} with a new run time $\tau_R'$. For Levy swimmers, the runtime is sampled from Pareto distribution $\phi(\tau)=(\alpha\tau_0^\alpha)/(\tau+\tau_0 )^{\alpha+1}$, where the parameter $1<\alpha<2$ controls the power-law index \cite{RN21}. In all simulations, translational thermal noise \revise{was} chosen such that the passive translational diffusivity is $0.1 \mu m^2 /s$. 
Bacteria shape \revise{was} simplified as spheres with negligible size. 
For the mechanism demonstration in Fig.2J, we simulate 1,000,000  particles with a persistent run time $\tau_R=2 s$ for 200 s in a 2D channel $50 \mu m$ wide. \revise{A} periodic boundary condition is always imposed along the \revise{direction of the} channel. As a result, the channel is effectively infinitely long, and the obstacles are also repeated every 100\revise{$\mu$m}. For the designed channels, sliding (for the particle dynamics) and no-slip (for the fluid dynamics) boundary conditions are imposed at the geometric boundary of the walls, except for the surface coating case where the no-slip boundary is at the wall and the sliding boundary condition for the particles are set at 3 $\mu m$ away from the wall. 

\paragraph{Geo-FNO model and machine learning setup}
The catheter design problem is an \revise{Stochastic Partial Differential Equation} (SPDE) constrained optimization problem, where the objective function $\langle x_{up} \rangle$ depends on the SPDE solution of the coupled fluid-particle problem. Traditional optimization approaches  require repeatedly evaluating such expensive computational models. And an adjoint solver is required when gradient-based optimization is applied. To overcome these computational challenges, we train\revise{ed} a Geo-FNO $\mathcal{G}$ as a surrogate model for the forward coupled fluid particle simulation that maps  the channel geometry to the bacteria population function $\mathcal{G}:c \to \rho$. In contrast, prior work using AI approaches for various design problems only  \revise{chose} a few parameters that are input to  traditional solvers of SPDE \cite{frazier2016bayesian, zhang2020bayesian}.  
The full model consists of 5 Fourier neural layers with the GeLU activations following and has a fast quasi-linear time complexity. 
We perform\revise{ed} coupled fluid-particle  simulations using both the ABP and Levy RTP models for three \revise{maximum} flow \revise{speeds} ($5, 10, 15 \mu m/s$) to generate training and testing data for the Geo-FNO. For the training data, we generate\revise{d} 1000 simulations in parallel (10 minutes per instance), with the design in each simulation randomly selected from the following  parameter space: obstacles with height $20 \mu m < h < 30 \mu m$ are periodically placed on the channel walls with inter-obstacle distance $60 \mu m < d < 250 \mu m$, the base length  satisfies $15 \mu m < L < d/4$, and the tip position satisfies $-d/4 < s < d/4$. The constraints on these parameters \revise{were} chosen to satisfy fabrication limits and physical conditions for the vortex generation mechanism (Fig.2 B,C, \revise{s}ee more discussions in Supplementary Information and Fig.S2). The dataset is stored to be reused for future tasks. We use\revise{d} the  relative empirical mean square error as the loss function. The model training \revise{took} 20 minutes with \revise{the }Adam \revise{optimizer} on an Nvidia GPU. It gets around 4\% relative error on the 100 testing data points.
\paragraph{Fast inverse design with gradient-based optimization}
The benefit of our AI approach is the speedup   compared to traditional solvers, and differentiability allows the use of fast gradient-based methods for geometry design optimization. Each evaluation takes only 0.005 seconds on GPUs in contrast to 10 minutes by using \revise{GPU-based} coupled fluid-particle simulations, and therefore it is affordable to do thousands of evaluations in the optimization procedure. Moreover, we use automatic differentiation tools of deep learning packages to efficiently compute gradients with respect to design variables enabling the use of gradient-based design optimization methods. 
During optimization, we start from initial design parameters ($d=100, h=25, s=10, L=20$) $\mu$m, and update them using  the BFGS algorithm to minimize the objective function $\langle x_{up} \rangle$ post-processed from the bacteria population predicted by Geo-FNO. When the optimization gets trapped in a local minimizer, the optimization restarts from an initial condition obtained by perturbing the recorded-global minimizer with a random Gaussian noise sampled from $\mathcal{N}(0, I)$. The proposed randomized BFGS algorithm guarantees the recorded-global minimizer monotonically decreases.  The optimization loss trajectory is depicted in Supplementary Fig.S1, which is reduced from $L=6.68 \times 10^5$ to  $L=2.18 \times 10^5$.
The AI-based optimization took about 1500 iterations to find the optimal design (15 seconds in total), whereas the traditional genetic optimization \revise{took} three days with the numerical solver. 
Within imposed parameter constraints, $\langle x_{up} \rangle$ is neither convex nor monotonic with respect to these design variables, but is generally smaller with larger $h$, smaller $d$, and larger $s$ (Fig.~S3).  
The final optimized design is with ($d=62.26,h=30.0,s=-19.56,L=15.27$)  $\mu$m.

\paragraph{Bacterial strains, culture conditions, materials, and chemicals}
We use\revise{d} wild-type BW25113 \revise{{\it E. coli}} with kanamycin resistance for the 3D catheter long-term experiment and BW25113 \revise{{\it E. coli}} expressing mScarlet red fluorescent protein with kanamycin resistance for the microfluidic experiments. A single colony of the bacterium of interest was picked from a freshly streaked plate and suspended in LB medium to create a bacterial inoculum. The start\revise{ing} culture \revise{was} cultured overnight at 37°C in LB medium to achieve a final concentration of approximately OD600 0.4. For the microfluidic experiments, 300\revise{~$\mu L$} of the start\revise{ting} culture is transferred to a new flask with 100 mL LB median and cultured at 16°C until OD600 reaches 0.1-0.2. \revise{B}acteria are washed twice by centrifugation (2300g for 15 min), and the cells were suspended in a motility imaging medium composed of 10 mM potassium phosphate (pH 7.0), 0.1 mM K-EDTA, 34 mM K-acetate, 20 mM sodium lactate, and 0.005\% polyvinylpyrrolidone (PVP-40) \cite{RN15}. The use of this medium allows for the preservation of bacterial motility while inhibiting cellular division. The final concentration of the bacteria in the reservoir \revise{has} OD600 at 0.02. For the 3D catheter long-term experiments, 3 mL of the starter culture is transferred to a new flask with 500 mL LB median and cultured at 16°C until OD600 reaches 0.4. The bacteria are directly used and injected into the bacteria reservoir. Kanamycin \revise{was} added to all the culture median and LB plates. The mobility of the bacteria \revise{was} checked under the fluoresce microscope 10 min before the experiment (\revise{observed} under DIC for BW25113  and RFP for \revise{the} BW25113 mScarlet strain). 

\paragraph{Microfluidic Experiments}
To demonstrate the mechanism of our design and test the effectiveness of the optimized structure, we fabricate\revise{d} quasi-2D micro-fluidic channels to observe bacteria motion under a microscope. These microfluidic devices \revise{were} fabricated using photolithography and PDMS soft-lithography. As shown in the schematic of Fig.~3A, one end of the microfluidic channel connects to a syringe filled with imaging solution, and the other end connects to a reservoir of \revise{{\it E. coli}}. The flow rate is controlled by tuning the height of the syringe with respect to the outlet downstream. Fluorescent beads \revise{were} injected into the imaging solution as passive tracers to monitor the flow rate in real time. The high-speed video was achieved using an Olympus BX51WI microscope with two Photometrics Prime95B cameras connected using a W-View Gemini-2 Optical Splitter from Hamamatsu. An Olympus 20X dry objective lens was used. Time-lapse images were acquired at ~12.4 frames per second with 488 \revise{nm} laser intensity set at 20\%. The microscope’s focal plane \revise{was} fixed near the middle of the channel in the depth z-direction to avoid recording bacteria crawling on the top and bottom sides of the channel. Experiments were performed on three different days with independent batches of \revise{{\it E. coli}}. cultures, with five 15-minute recordings each day.  ImageJ software (Fiji) was used for video post-processing \cite{RN37,RN38} to extract the trajectories of the bacteria. The trajectories are filtered by their linearity of forward progression to eliminate the fast-moving downstream ones and visually highlight the upstream swimming ones. We estimate the time interval for the upstream swimming to be 10 s before the fall-off. The maximum flow \revise{speed} is defined as the highest flow speed along the channel’s centerline. The instantaneous maximum flow rate is estimated by averaging the fastest velocities of bacteria and fluorescent beads along the centerline during the upstream-fall-off interval. Several video recordings are provided in the supplementary materials. 

\paragraph{3D Catheter Long-time Experiments}
Prototype catheter tubes (both geometric designs and smooth tubes) were printed using a Connex-Triplex 3D printer. The inside of the tubes with the designed obstacles are similar to the quasi-2D structures but enlarged and revolved about the center line of the channel so that the obstacles are extruded rings on the inner walls. Considering the 3D printing accuracy available and the scale of typical catheters, these prototypes are 1.6 cm in inner diameter. For the designed tubes, the spacing between the extruded rings is 1 mm. For ease of clearing the supporting material from 3D printing, each tube is printed as two halves with a tenon shape along the long side and assembled into a complete tube after removing the supporting material. As shown in Fig.5A, the top end of the tube is connected to a syringe controlled by a mechanical pump that keeps a constant flow rate. The bottom end of the tube is connected to an 80 mm diameter petri dish as a reservoir for \revise{{\it E. coli}}. After 1 hr, the tube is cut into 2 cm long segments, and the liquid contained inside each segment is transferred to a culturing plate, discarding the most upstream and downstream segments. After culturing the plates for 24 hrs at room temperature, the number of bacteria colonies in each plate is counted to reflect the amount of contamination in the corresponding part of the tube. We selected four circular, equally distant regions of 8 mm diameter in plate to count the number of colonies in these four areas (Fig.S5). The total number of colonies in the plate is estimated by multiplying the total number in the four circles by the area ratio 25, of the entire plate v.s. the four areas. We denote the total number in the plate to be 30,000 when too many colonies are on the plate and they become too crowded/overlapping to count precisely.

%% file: supp.tex
\renewcommand\thefigure{S\arabic{figure}} 
\setcounter{figure}{0}

\section*{Supplementary Discussion}

\paragraph{Example bacteria trajectories under different geometric and flow conditions}
In Fig.~S1 we show a few more example trajectories of Brownian bacteria. Comparing Fig.~S1(A) with (C), without flow the asymmetric obstacle shape provides an additional rectification effect. Comparing Fig.~S1(A)(C) to (B)(D), the flow vorticity provides a significant additional disruption to redirect the bacteria. 

\begin{figure}[h]%
\centering
\includegraphics[width=0.7\textwidth]{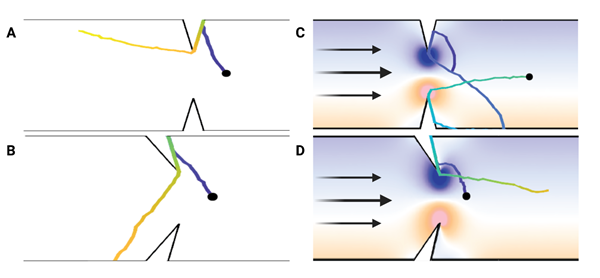}
\caption{Example simulated trajectories of active Brownian particles in channels with (A)(C) symmetric obstacles and (B)(D) asymmetric obstacles. (A)(B) without fluid flow. (C)(D) with the flow.}
\label{figS1}
\end{figure}

\paragraph{Fluid vortices around the tip of obstacles}

In Fig.~S2 we show how the relative size and spacing of obstacles affect the effective vorticity of the fluid flow. To make fair comparisons, we measure the maximum vorticity  and then normalize it by the effective flow speed  
\begin{equation}
\Tilde{\Omega} = \frac{a\Omega}{U_0},
\end{equation}
where $a$ is the particle diameter.
The effective flow speed is measured as the minimum value along the center line of the channel, with the channel width W.

Fig.~S2 (A)(B) compare the limits of small and large inter-obstacle distance $d$. In the limit of very large separation between neighboring obstacles (Fig.~S2 B), the vortices are independent of each other and saturate to a constant as d/W increases (Fig.~S2 G). However, when the vortices are brought very close to each other (Fig.~S2 A), they start to interfere and the resulting effective vorticity is reduced as d decreases (Fig.~S2 G). Hence there exists a threshold value for the inter-obstacle distance such that the vortices do not overlap and fully develop. This condition is beneficial to the suppression mechanism as stronger vortices can disrupt the upstream swimming trajectories more. 

Fig.~S2 (C-F) show the enhancement of effective vorticity as the obstacle height h increases. This is due to two reasons: (1) The flow lines are ‘pinched’ when passing over an obstacle, such that the flow speed at the obstacle tip is enhanced, which is proportional to the vorticity. (2) The more acute angle of the tip (a larger effective local curvature) induces a stronger ‘bending’ of the flow lines. However, the height should not be arbitrarily large: the pressure drop will increase significantly as a larger h indicates a smaller effective channel width, \revise{as shown in Fig.~S2(F)}. In the limit of h=W/2, the channel is completely blocked by the obstacles and the pressure drop required to flow diverges.  

Considering all these effects, together with the fabrication accuracy of microfluidics, we formulated the constraint optimization problem as described in the main text and Materials and Methods. 

\begin{figure}[htb]%
\centering
\includegraphics[width=0.7\textwidth]{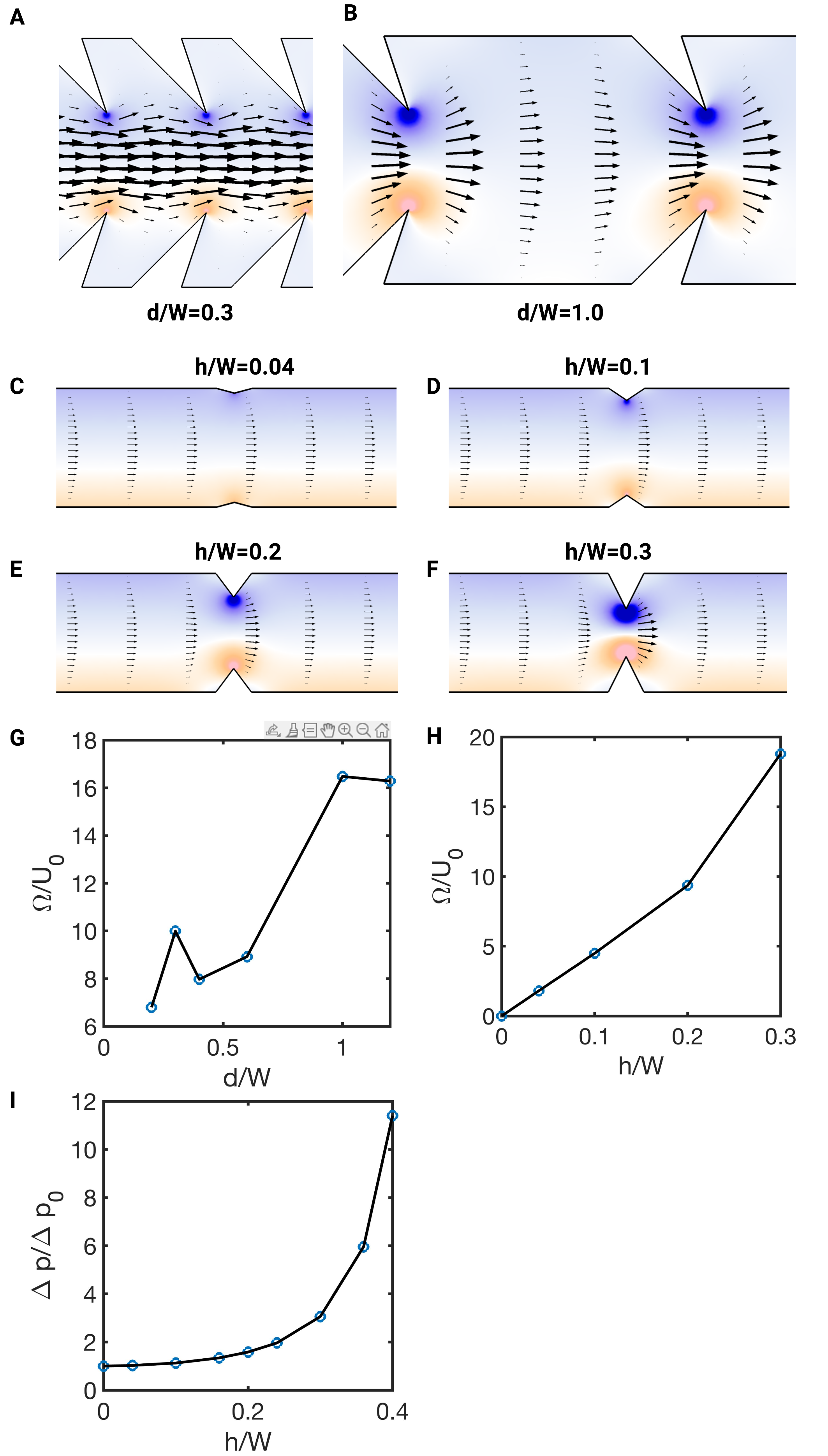}
\caption{(A-H)Normalized vorticity as a function of the obstacle height $h$ and the distance between obstacles $d$. \revise{(I) Normalized pressure drop over one period along the channel as a function of the normalized obstacle height $h/W$.}}
\label{figS2}
\end{figure} 

\paragraph{Discussion on Geo-FNO aided design optimization}

Anti-infection catheter design is essentially a partial differential equation constraint optimization problem.  However, solving these design optimization problems requires repeatedly evaluating computational models to explore the design space. \revise{Hence}, the present coupled fluid particle simulation, which takes about 10 minutes, is expensive or even unaffordable for this purpose. Moreover, design optimization generally requires an adjoint solver to compute the gradient for efficient optimization algorithms, but the randomness introduced by the Levy RTP model makes it impractical to implement. All these challenges motivate\revise{d} us to consider machine learning-based surrogate models, Geo-FNO, to accelerate the design optimization process.

We train the Geo-FNO model on 1000 simulations uniformly generated from our design space, and test it on 100 randomly generated designs. 
The model takes the shape of the channel as the input, and outputs the bacteria density as a 1D function. 
The training error and test error are depicted in Fig.~S3, no overfitting is observed, and the average relative test error is about 4$\%$.  After training, Each evaluation of the map from the channel geometry to the bacterial population takes only 0.005 seconds on GPUs in contrast to 10 minutes by using coupled fluid-particle simulations, and therefore it is affordable to do thousands of evaluations in the optimization procedure. 

For the optimization, the forward map takes these four design parameters d, L, h, s, generates channel geometry, predicts the bacteria population with Geo-FNO, and finally computes the objective function $\revise{\langle x}_{up}\rangle$. Automatic differentiation tools embedded in the deep learning package (i.e., Pytorch)  
are used to  efficiently compute gradients with respect to design variables enabling the use of gradient-based design optimization methods. 

We start from initial design parameters $(d=100, h=25, s=10, L=20)~\mu m$, and update them using  the BFGS algorithm with Strong Wolfe line search to minimize the objective function $\langle \revise{x}_{up} \rangle$. To enforce the constraints about these design parameters, exponential transforms are applied to the design parameters. For example to enforce $x_{min} \leq x \leq x_{max}$, $x$ is defined as $x=\phi(\theta) = x_{min} + (x_{max} - x_{min})/(1+e^\theta)$. This ensures that Geo-FNO remains in the interpolation regime and the final design satisfies manufacturing conditions.

\begin{figure}[htb]%
\centering
\includegraphics[width=0.7\textwidth]{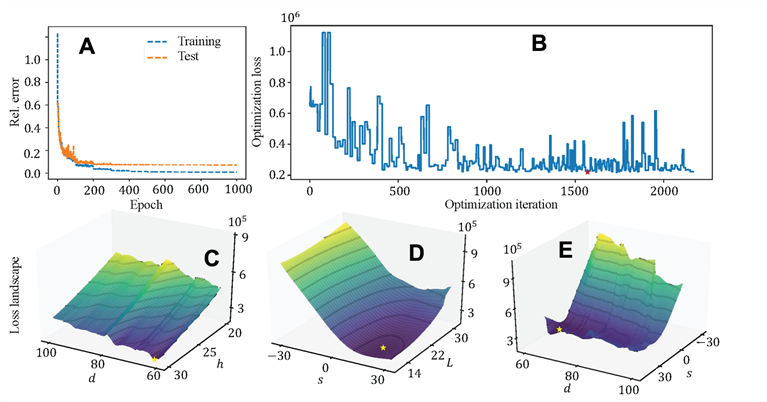}
\caption{Optimization. A.Training and test errors for Geo-FNO, both errors converge without overfitting; B.Optimization loss obtained by the randomized BFGS algorithm accelerated by our Geo-FNO surrogate model, the recorded-global loss is obtained at about 1500 iterations. C.Visualization of the loss landscape around the optimized design at the d-h cross-section obtained by Geo-FNO; D. Visualization of the loss landscape around the optimized design at the s-L cross section obtained by Geo-FNO; E. Visualization of the loss landscape around the optimized design at the d-s cross-section obtained by Geo-FNO. }
\label{figS3}
\end{figure}

Another challenge is related to local minimizers, since most partial differential constraint optimization is non-convex. When the optimization gets trapped in a local minimizer, the optimization restarts from an initial condition obtained by perturbing the recorded-global minimizer with a random Gaussian noise sampled from $N(0, I)$. The optimization loss vs. Optimization iteration curve is depicted in Fig.~\ref{figS3}B. The recorded-global minimizer is obtained at about 1500 iterations, the loss is reduced from $\revise{\langle x_{up}\rangle}=6.68 \times  10^5$ to   $\revise{\langle x_{up}\rangle}=2.18 \times 10^5$. Several cross-sections of the loss function landscape around the final optimized design are presented in Fig.~\ref{figS3}C-E, Within imposed parameter constraints,   the landscape near the optimized design is neither convex nor monotonic with respect to these design variables, but the loss is generally smaller with larger $h$, larger $s$, smaller $d$, which indicates the channel design is more effective when the height of the obstacle is large, the tip points towards downstream, and obstacle is more frequent.

\paragraph{Bacteria supercontamination in all flow rates in 3D macroscopic experiments}
In Fig.~S4 we show the supercontamination measured in all 3 flow rates.

\begin{figure}[htb]%
\centering
\includegraphics[width=0.7\textwidth]{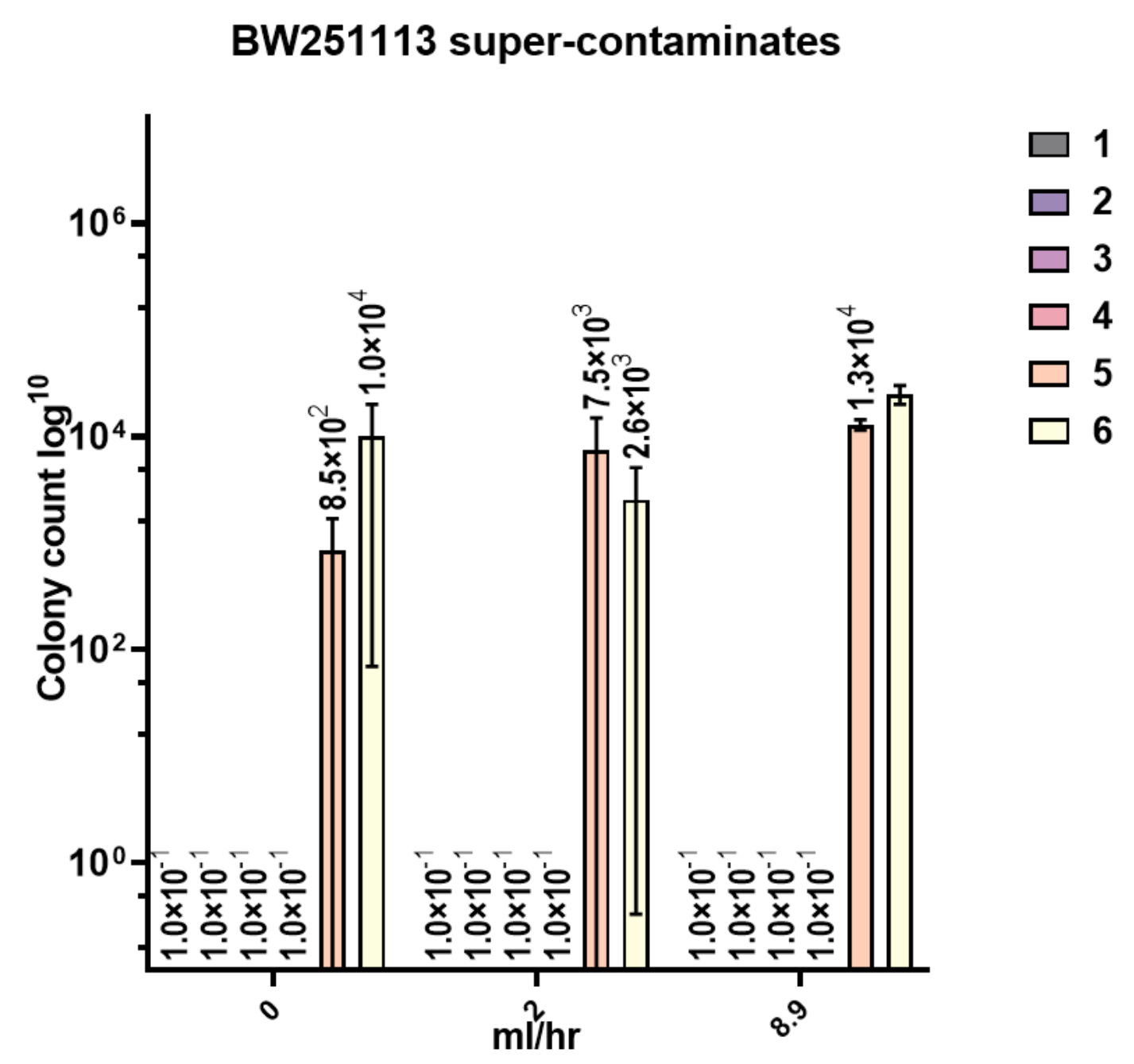}
\caption{
Supercontamination measured in all 3 flow rates.}
\label{figS4}
\end{figure}

\section*{Supplementary Methods}

\paragraph{Bacterial colony counting method} 

\begin{figure}[htb]%
\centering
\includegraphics[width=0.5\textwidth]{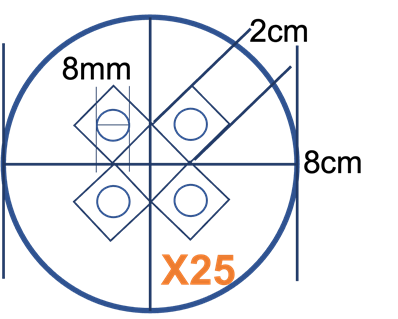}
\caption{Schematic of bacterial colony counting method on culturing plates.}
\label{figS5}
\end{figure}

In the macroscopic experiments, we count\revise{ed} the number of bacterial colonies on the plates after curing them for 24 hrs at room temperature. Four round areas of equal size (8mm diameter) \revise{were} selected, as shown in Fig.~\ref{figS5}, and only the colonies within these areas are counted. \revise{We then} multiplied the result by the ratio of the total plate area (8cm diameter) / counting areas = 25 to estimate the total number of colonies. When too many colonies are present on a plate, it becomes impossible to distinguish them \revise{and} we denote\revise{d} the number of colonies as an arbitrarily large number 30000.

